\documentstyle[preprint,aps,version2]{revtex}
\draft
\tightenlines
\def\Journal#1#2#3#4{{#1} {\bf#2}, {#3} {(#4)}}

\def\NP{{ Nucl. Phys.} }
\def\PLB{{ Phys. Lett.}  B}
\def\PRL{ Phys. Rev. Lett.}
\def\PRD{{ Phys. Rev.} D}
\def\ZPC{{Z. Phys.} C}
\def\EPJC{{Eur. Phys. J.} C}

\def\be{\begin{equation}}
\def\ee{\end{equation}}
\def\bea{\begin{eqnarray}}
\def\eea{\end{eqnarray}}

\def\qbar{{\bar q}}
\def\ubar{{\bar u}}
\def\dbar{{\bar d}}
\def\sbar{{\bar s}}

\def\NP{{ Nucl. Phys.} }

\begin{document}
\begin{titlepage} 
%\begin{flushright}
%BIHEP-TH-96-30 \\
%November 1996
%\end{flushright}

\bigskip
\begin{title}
{\large\bf On the phenomenological analyses of $s$-$\sbar$ asymmetry in the nucleon sea}

\end{title}
\author{Fu-Guang Cao\thanks{E-mail address: f.g.cao@massey.ac.nz.}
 and A. I. Signal\thanks{E-mail address: a.i.signal@massey.ac.nz.}}
\begin{instit}
Institute of Fundamental Sciences, Massey University,
Private Bag 11 222, Palmerston North,
New Zealand
%\thanks{E-mail address: f.g.cao@massey.ac.nz; a.i.signal@massey.ac.nz}
\end{instit}

\begin{abstract}

%Meson cloud model (MCM) and light-cone model (LCM)
Two phenomenological models
which give opposite predictions for the $s$-$\sbar$ asymmetry
in the nucleon sea are re-analyzed carefully.
It is pointed out that although the quantitative results
in both models depend dramatically on the parameters,
the predictions for the shape of $s(x)-{\bar s}(x)$
in the two models are parameter independent and opposite.
Thereby the coming experiments are likely to be able to distinguish the two models. 
We find that the reason for the two models giving opposite predictions
%for the shape of $s(x)-{\bar s}(x)$
is that the fluctuation functions and parametrizations for the strange (anti-strange)
quark distribution in the baryon (meson) in the two models are quite different.
To further investigate these models, we use
the same parametrizations for the strange (anti-strange) distributions
of the baryon (meson) in the two models.
We find that one of the models
depends strongly on the parameter which controls
the behavior of the meson-baryon fluctuation function.
Also the two models agree on the shape and size of $s$-$\sbar$
for some values of the model parameters,
but can disagree strongly for others.
%Consequently, the prediction for the $s$-${\bar s}$ asymmetry
%from the MLCM can be dramatically different from that from MCM
%in some parameter ranges
%as well as very similar in the other parameter space.
%The coming experimental data will examine these calculations,
%or equally, provide information on both the fluctuation function
%and strange and anti-strange distributions in the $\Lambda$ ($\Sigma$) baryon
%and $K$ meson.

\bigskip
\noindent
PACS number(s): 13.60.Hb; 11.30.Hv; 12.39.Ki; 13.88.+e
\end{abstract}
\end{titlepage}

\section{Introduction}
Studying the light quark content in the nucleon sea is important to the
understanding of nucleon structure as well as strong interaction.
There is strong experimental evidence that the light quark sea
is flavor asymmetric i.e. ${\bar u}\not= {\bar d}$ \cite{New,NA51,E866,Hermes}.
%especially after E866/NuSea Collaboration publishes
%its measurement on the $\dbar$ and $\ubar$ content in the proton \cite{E866}.
Another interesting question concerning the dynamics of the light quark
sea of the nucleon is quark-antiquark ($q$-${\bar q}$) asymmetry.
The nucleon sea can be broken down into perturbative (``extrinsic")
and non-perturbative (``intrinsic") parts.
The perturbative sea is created from gluon-splitting
and can be calculated from perturbative QCD
as the $q \qbar$ pair exists only for short times.
In the leading twist approximation, the perturbative sea is symmetric,
i.e. $q={\bar q}$.
The non-perturbative sea, however, may exist over a long time and it has a strong
connection with the ``bare" nucleon.
There is no fundamental theoretical principle and/or experimental evidence
which demands that the non-perturbative sea is symmetric.
Although it is usually assumed that the quark sea in the nucleon is equal to
the anti-quark sea,
one should note that the $q$-$\qbar$ symmetry may be violated to some extent.
Because one can not distinguish the sea up and down quarks
from the valance up and down quarks in the nucleon bound state,
it is difficult to study the $q$-$\qbar$ asymmetry of the up quark sea
and the down quark sea in experiments.
%The most possible scheme is to study the strange content of the nucleon sea.
However, the strange content of the nucleon sea is accessible to experiments
\cite{Boros,CCFR,NMC}.
Although there is still large uncertainty in the extraction of $s$ and ${\bar s}$
distributions from experimental data \cite{Boros,CCFR,NMC},
the analysis of \cite{NMC} strongly suggests that $s(x) \not= \sbar(x)$.
We can expect that the experimental data will be improved in the near future.

There have also been some theoretical analyses on this issue
\cite{Burckardt,Signal,Holtmann,Ma,Christiansen}.
From the chiral Gross-Neveu model, Burckardt and Warr \cite{Burckardt}
suggested that a large $s$-${\bar s}$ asymmetry may exist in the nucleon sea.
Due to its success in the study of the flavor asymmetry of the nucleon sea,
the meson cloud model (MCM) has also been used in the study of the strange sea
of the nucleon.
Employing the meson cloud model with the fluctuation function calculated
from covariant perturbative theory and cloudy bag model \cite{bagmodel}
Signal and Thomas \cite{Signal} predicted that
the $s$ and ${\bar s}$ can have quite different shapes, although the quantitative
results depended on the bag radius.
Holtmann, Szczurek and Speth \cite{Holtmann}
performed their analysis using the meson cloud
model with the fluctuation function being calculated from time-order perturbative
theory in the infinite momentum frame and the parameter in the
form factor being fixed by the high-energy particle production data.
It was found that $s < {\bar s}$ in small $x$ region and $s > {\bar s}$
in large $x$ region.
Brodsky and Ma \cite{Ma} proposed a light-cone two-body wave function
model (LCM) for the description of the meson-baryon fluctuation, and
they obtained a significantly different conclusion from \cite{Holtmann}:
$s > {\bar s}$ ($s < {\bar s}$) in the small (large) $x$ region.
The quantitative predictions in \cite{Ma} dramatically depend on the normalization
of the fluctuation.
More recently, Christiansen and Magnin \cite{Christiansen} arrived a similar
conclusion as \cite{Ma} by employing both effective and perturbative
degrees of freedom and incorporating the recombination mechanism which
has been well used in the study of the hadron production.
It is worth to note that
the same physical picture -- nucleon fluctuating to meson and baryon,
has been used in both MCM \cite{Signal,Holtmann} and LCM \cite{Ma},
but the predictions for the $s(x)-\sbar(x)$ are quite different.

The purpose of this paper is to re-analyze the $s$-${\bar s}$ asymmetry
of the nucleon sea in the frameworks of both
the MCM \cite{Signal,Holtmann} and LCM \cite{Ma}, and to find the reason
that quite different predictions are obtained from the two models
in which the same physical picture is employed.
In sections II and III, we analyze the meson cloud model and
light-cone model respectively.
In section IV, we investigate the model-dependence of the two models
and present more discussions.
The last section is reserved for a summary.

\section{Meson cloud model}

The meson cloud model was first suggested and developed in the studies of
low energy physics and it has been proved to be a successful tools in
understanding both nucleon structure and dynamics.
Lately, this model has been applied to
%the studies of high energy physics especially in
the studies of structure function and sea content
of nucleon \cite{Sullivan,Thomas83}.
The basic idea of the meson cloud model is that the nucleon
can be viewed as a bare nucleon surrounded by a mesonic cloud.
The nucleon wave function is expressed in terms of a series of
baryon and meson components.
For the strange content of the nucleon sea, the important components
are $\Lambda K$ and $\Sigma K$ Fock states,
\bea
|N\rangle_{\rm physical} = |N\rangle_{\rm bare}
&+&\sum_{\lambda \lambda^\prime} 
\int dy \, d^2 {\bf k}_\perp \, \phi^{\lambda \lambda^\prime}_{\Lambda K}(y,k_\perp^2)
\, |\Lambda^\lambda(y, {\bf k}_\perp); K^{\lambda^\prime}(1-y,-{\bf k}_\perp)
\rangle \nonumber \\
&+&\sum_{\lambda \lambda^\prime} 
\int dy \, d^2 {\bf k}_\perp \, \phi^{\lambda \lambda^\prime}_{\Sigma K}(y,k_\perp^2)
\, |\Sigma^\lambda(y, {\bf k}_\perp); K^{\lambda^\prime}(1-y,-{\bf k}_\perp)
\rangle \nonumber \\
&+&\cdots,
\label{NMCM}
\eea
where $\phi^{\lambda \lambda^\prime}_{\Lambda K(\Sigma K)}(y,k_\perp^2)$ 
is the wave function
of the Fock state containing a $\Lambda$ ($\Sigma$) baryon
with longitudinal momentum fraction $y$, transverse momentum ${\bf k}_\perp$,
and helicity $\lambda$,
and a $K$ meson with momentum fraction $1-y$, transverse momentum $-{\bf k}_\perp$,
and helicity $\lambda^\prime$.
The model assumes that the lifetime of a virtual baryon-meson Fock state is much larger
than the interaction time between the hard photon and nucleon in deep inelastic scattering,
thus the non-perturbative contributions to the strange and anti-strange
distributions in the proton, $s^N$ and $\sbar^N$, can be written as convolutions
\bea
s^N (x)=\int^1_x \frac{dy}{y} f_{B K}(y) s^B(\frac{x}{y}),
\label{sN}
\eea
\bea
\sbar^N(x)=\int^1_x \frac{dy}{y} f_{K B}(y) \sbar^{K}(\frac{x}{y}),
\label{sbarN}
\eea
where $B=\Lambda \, (\Sigma)$,
$s^B$ and $\sbar^{K}$ are the $s$ and $\sbar$ distributions in the
$\Lambda$ ($\Sigma$) and $K^+$ respectively,
and $f_{B K}$ is fluctuation function which describes the possibility for a nucleon
fluctuating into a $\Lambda K$ ($\Sigma K$) state,
\bea
f_{BK}(y)=\sum_{\lambda \lambda^\prime}\int^\infty_0 d k_\perp^2
\left | \phi^{\lambda \lambda^\prime}_{B K}(y, k_\perp^2)\right |^2.
\label{fMCM}
\eea
From Eqs.~(\ref{sN}) and (\ref{sbarN}) we know that
$s$ and $\sbar$ distributions in the nucleon are different
and the difference $s - \sbar$ depends on both the fluctuation
functions ($f_{BK}$ and $f_{KB}$) and valance quark distributions
in the baryon and meson ($s^B$ and $\sbar^K$).
Due to the baryons $\Lambda$ and $\Sigma$ being heavier than
the $K$ meson, the $f_{BK}(y)$ peaks at $y > 0.5$ while
the $f_{KB}(y)$ peaks at $y < 0.5$ (see Fig.~5), which suggests that
$s^N > \sbar^N$ in the large $x$ region.
On the other hand, in the large $x$ region the $\sbar$ distribution of
the $K$ meson ($\sbar^K(x)$) is
generally believed to be larger than the $s$ distribution of
the baryon ($s^B(x)$) as the baryon contains one more valance quark
than the meson (see Fig.~4).
This implies that  $s^N < \sbar^N$ in the large $x$ region.
The final prediction of the $x$ dependence of $s$-$\sbar$ asymmetry
will depend on these two competing effects.

From the consideration of momentum and charge conservation, we have
\bea
f_{BK}(y)=f_{K B}(1-y).
\label{relation}
\eea
It has been pointed out \cite{Holtmann,Zoller,Melnitchouk}
that the constraint Eq.~(\ref{relation}) can be guaranteed in the calculation
employing time-ordered perturbative theory (TOPT) in the
infinite momentum frame while it cannot be
satisfied automatically in the covariant perturbation calculation.
Another advantage of employing the TOPT in the
infinite momentum frame is that the intermediate particles (baryons and messons)
are on their mass-shell and so there is no
ambiguity associated with the possible off-mass-shell behavior of
their structure functions which are encountered in the covariant
perturbative formulations.

The wave function $\phi_{B K}^{\lambda \lambda^\prime}$ in Eq.~(\ref{fMCM}),
and thereby the fluctuation function $f_{B K}(y)$, can be calculated from
the effective meson-baryon-nucleon interaction Lagrangian,
\bea
L=g \cdot i {\bar \psi} \gamma_5 \phi \psi,
\eea
where $g$ is the effective coupling constant, and $\psi$ and $\phi$ are
the nucleon and pseudoscalar fields respectively.
Employing time-ordered perturbative theory in the infinite momentum frame,
%in which the relation Eq.~(\ref{relation}) can be guaranteed \cite{Holtmann},
we obtain
\bea
f_{B K}^{\rm MCM}(y)=\frac{g^2_{N B K}}{16 \pi^2}
\int^\infty_0 \frac{d k_\perp^2}{y(1-y)}
\frac{G^2_{B K}(y, k_\perp^2)}{(m_N^2-m_{B K}^2)^2}
\frac{(y m_N - m_B)^2+k_\perp^2}{y},
\label{fBKMCM}
\eea
where $m_{B K}^2$ is the invariant mass squared of the $\Lambda K$ ($\Sigma K$)
Fock state,
\bea
m_{BK}^2=\frac{m_B^2+k_\perp^2}{y}
+\frac{m_K^2+k_\perp^2}{1-y},
\label{mBK}
\eea
and $G_{BK}(y,k_\perp^2)$ is a phenomenological vertex form factor
for which we adopt a exponential form \cite{Holtmann}
\bea
G_{BK}(y,k_\perp^2)={\rm exp}
\left[\frac{m_N^2-m_{B K}^2(y,k_\perp^2)}{2\Lambda^2}\right].
\label{FF}
\eea
%The $(ym_N-m_B)^2$ and $k_\perp^2$ terms in Eq.~(\ref{fBKMCM}) come from
%the fluctuation processes with helicities $1/2 \ra (1/2,\, 0)$ and $1/2 \ra (-1/2, \, 0)$
%respectively. Thus they belong to the S-wave and P-wave parts of the nucleon
%wave function respectively.
It has been argued in Ref.~\cite{Holtmann}
from studies of baryon production processes
that a unique cut-off parameter $\Lambda=1.08$ GeV can be used
for all vertices involving octet baryons and pseudoscalar or vector mesons.
The fluctuation functions for the different charge states
can be obtained by using the following relations:
\bea
f_{\Lambda K^+}(y) &=& f_{\Lambda K}(y),
\label{fLambdaK} \\
f_{\Sigma^+K^0}(y) &=& 2 \, f_{\Sigma^0K^+}(y) = 2 \, f_{\Sigma K}(y),
\label{fSigmaK}
\eea
where $f_{\Lambda K}(y)$ and $f_{\Sigma K}(y)$ are given by Eq.~(\ref{fBKMCM})
with $B$ being $\Lambda$ and $\Sigma$ respectively.
We take the effective coupling constants $g^2_{N \Lambda K}/4\pi=13.7$
and $g^2_{N \Sigma K}/4\pi=3.7$ \cite{Signal,Aubert}.

For the $s$ distribution in the $\Lambda$ ($\Sigma$) it is common practice
to use the parametrization for the valance quark distribution in the nucleon
via relation \cite{Signal,Holtmann,Christiansen}
\bea
s^B=\frac{u^N}{2}.
\label{sB}
\eea
Here, we adopt the next-leading-order parametrization given in
\cite{GRV98} (GRV98) for $u^N$ at scale $\mu_{NLO}^2=0.40$ GeV$^2$,
\bea
u^N(x,\mu_{NLO}^2)=0.632 x^{-0.57} (1-x)^{3.09} (1+18.2 x).
\label{uN}
\eea
For the $\sbar$ distribution in the $K$, we adopt the parametrization
of \cite{GRS98} (GRS98) which is obtained by connecting ${\bar s}^K$ with
the valance quark distribution in the pionic meson
based on the consistent quark model,
\bea
\sbar^{K^+}(x,\mu_{NLO}^2)=v^\pi(x,\mu_{NLO}^2)-u^{K^+}(x,\mu_{NLO}^2),
\label{sbarK}
\eea
with
\bea
v^\pi(x,\mu_{NLO}^2) &=&1.052 x^{-0.495} (1 +0.357 \sqrt{x}) (1-x)^{0.365}, \\
u^{K^+}(x,\mu_{NLO}^2) &=& 0.540 (1-x)^{0.17} v^\pi(x,\mu_{NLO}^2),
\eea
at scale $\mu_{NLO}^2=0.34$ GeV$^2$.
The numerical result is given in Fig.~1.
It can be found that $s^N >  \sbar^N$ as $0.02 < x < 0.21$ and
$s^N < \sbar^N$ as $x > 0.21$.
%Also the contribution form the $\Sigma K$ Fock state is important.

\section{Light-cone model}

As is well known, the nucleon is built up from three valance quarks plus
$q\qbar$ pairs and gluons which compose the nucleon sea.
The light-cone (LC) formalism \cite{LCF} provides a
convenient framework for the relativistic description of hadrons
in terms of quark and gluon degrees of freedom.
Combined with perturbation theory, the light-cone formalism has
been applied to many exclusive processes with large momentum transfer.
In this formalism, the quantization is chosen at a particular
light-cone time $\tau=t+z$.
%and all components have positive light-cone
%momentum $k_i^+$ with the total  light-cone momentum of the bound state
%being given by $P^+=\sum_i k_i^+$.
%Hence the vacuum is very simple and is an exact eigenstate of the full 
%Hamiltonian $H_{\rm LC}$.
The hadronic wave function, which describes the hadronic 
composite state at a particular $\tau$, can be expressed in terms of 
a series of light-cone wave functions multiplied by the Fock states.
For example,
\bea
| {\rm p} \rangle=|uud>\psi_{{uud}/{\rm p}}
+ |uud g\rangle \psi_{uud g/{\rm p}}
+\sum_{ q \bar q} |uud q \bar q \rangle \psi_{uud q {\bar q}/{\rm p}}
+\cdots.
\label{pLC}
\eea
%The quark distribution $q(x,Q^2)$ of the nucleon measured in deep inelastic scattering
%is given by the sum of the squares of the light-cone wave functions \cite{Brodsky,Diehl}
%\bea
%q(x,Q^2)=\sum_N\sum_j\int[dx]_N[d^2{\bf k}_\perp]_N
%\delta(x_j-x_{B}) \left| \psi_{N/p}(x_i,{\bf k}_{\perp i})\right|^2,
%\label{qxLC}
%\eea
%where $x_B$ is the Bjorken variable, $x_j$ is the light-cone momentum fraction
%of the struck quark, and
%\bea
%[dx]_N&=&\prod^N_{i=1}dx_i\delta(1-\sum_i x_i), \\
%\left[d^2{\bf k}_\perp\right]_N&=&\frac{1}{(16 \pi^3)^{N-1}}\prod^N_i d^2{\bf k}_{\perp i}
%\delta^2\left(\sum_i {\bf k}_\perp\right).
%\eea

As an approximation, Brodsky and Ma \cite{Ma} suggested
that the nucleon wave function Eq.~(\ref{pLC})
could also be expressed as a sum of baryon-meson Fock states similar to Eq.~(\ref{NMCM}).
In principle, the predictions obtained by employing effective degrees of freedom
[Eq.~(\ref{NMCM})] should coincide with the results from employing
the quark-gluon degrees of freedom [Eq.~(\ref{pLC})].
The probability of the baryon-meson fluctuation should decrease with
the invariant mass of the baryon-meson Fock state increasing.
So the $\Lambda K$ and $\Sigma K$ Fock states are the most important states in
the study of the strange sea of nucleon, whereas the higher Fock states
%with the baryon and/or the meson being orbital excited
will be less important.
It was pointed out in Ref. \cite{Ma} that the possibility for finding the $\Lambda K^{* +}$
state is only about $5\% \sim 10\%$ of the possibility for finding the
$\Lambda K^+$ state.

In the light-cone baryon-meson fluctuation model,
the same two-level convolution mechanism and formula
(see Eqs.~(\ref{sN}), (\ref{sbarN}), (\ref{relation}) and discussion following them)
are employed to evaluate the non-perturbative contribution to
the $s$ and $\sbar$ content in the proton sea.
However, the two factors inside the convolution integrals,
the fluctuation function ($f_{BK}$ and $f_{KB}$) and valance quark distributions
in the baryon and meson ($s^B$ and $\sbar^K$), are now both
described using two-body light-cone wave functions.
 %In the light-cone baryon-meson fluctuation model,
The probability of the baryon-meson fluctuation (the fluctuation function)
%are modeled in the use of two-body wave function,
is given by
\bea
f_{BM}^{\rm LCM}(y)=\int_0^\infty \frac{d {\bf k}^2_\perp}{16 \pi^2}
		 \left| \psi(y,{\bf k}_\perp) \right|^2,
\label{fLCM}
\eea
where $\psi(y,{\bf k}_\perp)$ is a two-body wave function which is a function
of the invariant mass square of the baryon-meson state [see Eq.~(\ref{mBK})].
In Ref. \cite{Ma}, two wave function models, Gaussian type and power-law type,
were used, but nearly identical predictions for the strange content
of the nucleon sea were obtained.
Hence in our calculations, we adopt the Gaussian type wave function,
\bea
\psi(y,{\bf k}_\perp)=A\, {\rm exp}
\left[\frac{1}{8\alpha^2}\left(\frac{m_B^2+{\bf k}_\perp^2}{y}
+\frac{m_K^2+{\bf k}_\perp^2}{1-y} \right)\right],
\label{psi}
\eea
where $\alpha$ is a phenomenological parameter which determines the shape of
the fluctuation function and $\alpha=0.33$ GeV is taken in Ref. \cite{Ma}.
We will discuss the $\alpha$-dependence of the LCM in the next section.
Here we would like to point out that the constraint for the fluctuation
functions, $f_{BK}(y)=f_{KB}(1-y)$, is satisfied automatically in the LCM
[see Eqs.~(\ref{fLCM}) and (\ref{psi})].
For simplicity, the fluctuation function $f_{BM}^{\rm LCM}(y)$
was normalized to $1$ in Ref. \cite{Ma}, thus only strikingly different
$s$ and $\sbar$ distributions were observed 
but no absolute magnitude was given.
In order to make a meaningful comparison with the result of MCM,
we require that $f_{BK}^{\rm LCM}$ has the same normalization as
$f_{BK}^{\rm MCM}$ with $\Lambda=1.08$ GeV,
that is the probabilities of finding the $\Lambda K^+$
and $\Sigma^0 K^+$ Fock states
in the nucleon are $P_{\Lambda K^+}= 1.27\%$
and $P_{\Sigma^0 K^+}=0.25\%$
respectively\footnote{Note
from Eq.~(\ref{fSigmaK}) the probability of finding the $\Sigma^+ K^0$ Fock state
is $P_{\Sigma^+ K^0}=0.50\%$.}.
Furthermore, in Ref. \cite{Ma} the $s$ distribution in the $\Lambda$
and the $\sbar$ distribution in
the $K$ are also estimated by using a Gaussian type two body wave function
with the $ud$ pair in the $\Lambda$ being treated as a spectator,
\bea
s^B(x) &=& \int \frac{d {\bf k}_\perp^2}{ 16 \pi^2}
	    \left | A \,{\rm exp}\left[ \frac{1}{8 \alpha^2}
	    \left(\frac{m_s+{\bf k}_\perp^2}{x}
	    +\frac{m_D+{\bf k}_\perp^2}{1-x}\right)\right] \right |^2,
\label{sBMa} \\
\sbar^K(x) &=& \int \frac{d {\bf k}_\perp^2}{ 16 \pi^2}
	    \left | A \, {\rm exp}\left[ \frac{1}{8 \alpha^2}
	    \left(\frac{m_s+{\bf k}_\perp^2}{x}
	    +\frac{m_q+{\bf k}_\perp^2}{1-x}\right)\right] \right |^2.
\label{sbarKMa}
\eea
The mass parameters are taken as $m_q=330$ MeV for the light-flavor quark,
$m_s=480$ MeV for the $s$ quark,
and $m_D=600$ MeV for the spectator \cite{Ma}.
The strange and anti-strange distributions in the nucleon sea are obtained
via Eqs.~(\ref{sN}) and (\ref{sbarN}).
The numerical result is given in Fig.~1.
One can find that $s^N < \sbar^N $ as $0.02 < x < 0.24$
and $s^N > \sbar^N$ as $x > 0.24$,
which is opposite to the predictions from the MCM.

\section{Model-dependence and discussions}

The predictions from the two phenomenological models (MCM and LCM)
are quite different.
%Up to now, it is difficult to make a decision due to the lack of
%theoretical study from the first principle and experimental data.
%Both the two models should be treated as phenomenological models.
We want to make a comparison between the two models
and clarify the difference between them.
We then hope
that more precise experimental data will enable us to make a choice
between the two models, or at least determine their parameters.

The important parameters are $\Lambda$ in the MCM and $\alpha$ in the  LCM.
First, we study the effect of $\Lambda$ on the calculation by allowing
a range for $\Lambda$, $1.0$ GeV $ < \Lambda <1.2$ GeV.
It should be noted that the parameter $\Lambda$ can not be varied dramatically
since, in principle, it can be obtained by fitting the high-energy baryon production data.
($\Lambda=1.08 \pm 0.05$~GeV is given in \cite{Holtmann}.)
The numerical results are presented in Fig.~2.
It can be found that the quantitative results depend strongly on
the value of $\Lambda$,
which is not surprising since the value of $\Lambda$ governs the probability of
nucleon fluctuating to $\Lambda$ ($\Sigma$) baryon and $K$ meson.
(For $\Lambda=1.00$~GeV, $1.08$~GeV, and $1.20$~GeV,
$P_{\Lambda K}=0.83\%, \, 1.27\%,$ and $2.42\%$.) 
However, the prediction for the shape of $s(x)-\sbar(x)$ in the MCM is quite
independent of the value of $\Lambda$ --
$s(x)-\sbar(x)$ is negative for $0.02 < x <0.21$ and positive for $x > 0.21$
no matter which $\Lambda$ is chosen.
We note that our range of $\Lambda$ corresponds to a dipole form factor
with cut-off parameter in the range of $650\, - \, 850$~MeV \cite{PRD47},
which is fairly soft, and corresponds to a bag radius about $1$~fm
in the cloudy bag model.

In the LCM, the shapes of the fluctuation function [see Eq.~(\ref{fLCM})]
as well as the strange (anti-strange) distribution in the baryon (meson)
[see Eqs.~(\ref{sBMa}) and (\ref{sbarKMa})] are mainly determined by
the value of $\alpha$ -- they become broader as $\alpha$ increases.
The value of $\alpha$ reflects the strength of the interaction
potential in the two body bound state.
Phenomenological studies show that for the pion, a value of $\alpha$
in the range of $300 \, -\, 500$ MeV is favored \cite{alpha}.
However, there is little constraint on the corresponding parameter $\alpha$
when we consider the portion of the nucleon wave function arising from
one baryon and one meson Fock state,
since effective degrees of freedom are involved
and there are few studies employing this model.
In principle, the parameters $\alpha$ and $A$ involved in Eq.~(\ref{psi})
can be fixed by fitting high-energy baryon and meson production data
and experimental information on the $\dbar$-$\ubar$ asymmetry
in the proton \cite{New,NA51,E866,Hermes},
however this is beyond the scope of the present study and will be
the subject of future work \cite{CaoSignal2}.
%Here we would like to treat $\alpha$ as a phenomenological parameter
%whose value would be fixed by fitting the experimental data.
Due to the lack of information on the value of $\alpha$,
we study the $\alpha$ dependence of the calculation in the LCM
by varying $\alpha$ from $0.33$ GeV to $1.00$ GeV (see Fig.~3).
It can be found that the quantitative results strongly depend on the value of $\alpha$,
but the prediction for the shape of $s(x)-\sbar(x)$ stays the same and is opposite
to the prediction of MCM --
$s(x)-\sbar(x)$ is positive for $0.02 < x <0.24$ and negative for $x > 0.24$.
Fig.~3 is obtained by setting the probabilities of nucleon fluctuating to
$\Lambda K^+$ and $\Sigma^0 K^+$ Fock states to be $P_{\Lambda K^+}=1.27\%$
and $P_{\Sigma^0 K^+}=0.25\%$ respectively
which corresponds to taking $\Lambda=1.08$ GeV in the MCM.
Changing the probabilities $P_{\Lambda K}$ and $P_{\Sigma K}$
will change the prediction for $s(x)-\sbar(x)$ in magnitude,
but the predictions for the shape of $s(x)-\sbar(x)$ will persist.
Thus, we conclude that although the quantitative results in both models
depend on the parameters strongly,
the predictions for the shape of $s(x)-\sbar(x)$ in both models
are parameter independent.
It should be not difficult for the coming experimental data to examine
these two models since significantly different predictions are obtained
from both models.

It is interesting to note that although the same nucleon fluctuating to baryon
and $K$ meson physical picture and two level convolution formula
[see Eqs.~(\ref{sN}) and (\ref{sbarN})]
are employed in both MCM and LCM, the conclusion about the
$s$-$\sbar$ asymmetry from the two models are dramatically different.
We point out that both the strange and anti-strange distributions
in $\Lambda$ ($\Sigma$) and $K$, and the fluctuation functions
used in the two models, which are the two factors in the integrands of
the two-level convolution formulas [see. Eqs.~(\ref{sN}) and (\ref{sbarN})],
are quite different.
We compare the strange distribution employed in the two model in Fig.~4.
It can be seen that the strange distributions used in the LCM
exhibit a sharp peak in the medium $x$ region,
while the GRV98 and GRS98 parametrizations for $s^B$ and ${\bar s}^K$
emphasize the small-$x$ region strongly.
Also the strange (anti-strange) distributions employed in the LCM depends on
the parameter $\alpha$
-- they become less peaked as $\alpha$ increasing from $0.33$ GeV to $1.00$ GeV.
Up to now, experimental measurements and theoretical calculations from
first principles on the strange and anti-strange distribution in the baryon
$\Lambda$ ($\Sigma$) and meson $K$ are lacking
(but see Refs.~\cite{Alberg,Boros2}).
The realistic strange distribution may be different from the two-body wave function
prediction used in the LCM since only the lowest Fock state is considered
and the full result should be the sum of all Fock states \cite{Brodsky,Diehl}.
However, the $s^B(x)$ and ${\sbar}^K(x)$ obtained from the LCM being peaked at
$x < 0.5$ and $x > 0.5$ respectively reflects the fact that
the $s$ quark in the $\Lambda$ ($\Sigma$) should carry a smaller
amount of momentum than the $u d$ ($u u$) quark pair,
and the $\sbar$ quark in the $K^+$ should carry a large amount of momentum
than the $u$ quark.
The studies on the strange distribution in the baryon from a Nambu-Jona Lasinio
model \cite{Shigetani,Londergan}
and the anti-strange distribution in the $K$ meson from a Monte Carlo \cite{Edin}
support the above observation.
Thus, although it has been a common practice to employ the modified light valance
distributions of the nucleon and pion for $s^B$ and ${\sbar}^K$ respectively
here we would like to treat the strange distribution in the LCM and the MCM
as two phenomenological models.

In Fig.~5, we compare the fluctuation functions employed in the two models.
Although the same physical picture, nucleon fluctuating to baryon-meson Fock state,
has been adopted in both the meson cloud model and the light-cone model,
the suggested fluctuation functions are quite different
[see Eqs.~(\ref{fMCM}) and (\ref{fLCM})]: while a rather simple two-body
wave function is employed in the LCM\cite{Ma}, a more complex, effective
Lagrangian model is used in the MCM \cite{Signal,Holtmann}.
As has been pointed out in Sec.~III,
we adopt the same normalization condition for both $f_{BK}^{\rm MCM}$
and $f_{BK}^{\rm LCM}$ in order to make the comparison meaningful.
Using the same parameters given in Refs. \cite{Holtmann} and \cite{Ma},
that is, $\Lambda=1.08$ GeV in Eq.~(\ref{FF}) and $\alpha=0.33$ GeV in Eq.~(\ref{psi}),
we find that corresponding predictions for the fluctuation functions are
very different (see Fig.~5): $f_{BK}^{\rm MCM}(y)$ has a maximum value
at about $y=0.57$ while $f_{BK}^{\rm LCM}(y)$ has a maximum at about $y=0.67$.
In the non-relativistic limit, the ratio of the momentum fractions carried by
the $\Lambda$ and $K$ should be about
$y/(1-y) \sim m_\Lambda/m_K \simeq 2.3$, that is, the fluctuation function
should be sharply peaked at about $y=0.7$. Thus we know that the fluctuation function
obtained from the two-body wave function model is consistent with the non-relativistic
argument while the one obtained in the MCM is not.
Changing the parameters $\Lambda$ in Eq.~(\ref{FF}) and $\alpha$ in Eq.~(\ref{psi})
do affect the shapes of the fluctuation functions.
For example, decreasing $\Lambda$ and/or increasing $\alpha$ will decrease
the difference between the two models.
However, as we mention earlier,
there is only a small range of variation in $\Lambda$ allowed
from the consideration of the corresponding high-energy baryon production data.
($\Lambda=1.08 \pm 0.05$~GeV is given in \cite{Holtmann}.)
We present the results for $1.0$ GeV $ < \Lambda <1.2$ GeV in Fig.~6.
It can be found that the shape of fluctuation function
and the $y$-position at which the fluctuation function
exhibits a maximum value is insensitive to the value of $\Lambda$.
(The $y_{max}$ increases from $0.56$ to $0.58$ as $\Lambda$ decreases
from $1.20$ GeV to $1.00$ GeV.)
Consequently, the MCM's prediction for the shape of $s-\sbar$
is independent of $\Lambda$
although the quantitative result is sensitive to the value of $\Lambda$
as it has been shown in Sec. II (see Fig.~2). 
We study the $\alpha$ dependence of the fluctuation function
by taking $0.33$ GeV $ < \alpha <1.00$ GeV at given probabilities
$P_{\Lambda K^+}=1.27\%$ and $P_{\Sigma^0 K^+}=0.25\%$
which corresponds to take $\Lambda=1.08$ GeV in the MCM (see Fig.~5).
From Fig.~5, one can find that
the fluctuation function with $\alpha=0.33$ GeV is more peaked than
the fluctuation function with $\alpha=1.00$ GeV and
the $y_{max}$ changes from
$0.67$ to $0.58$ as $\alpha$ increases from $0.33$ GeV to $1.00$ GeV.
The lower value is close to the $y_{max}\simeq 0.57$ of $f_{BK}^{\rm MCM}$
with $\Lambda=1.08$ GeV.
Also the $f_{BK}^{\rm LCM}$ with $\alpha=1.00$ GeV has a similar symmetry
as the $f_{BK}^{\rm MCM}$.
Thus we expect that the prediction of the LCM with $\alpha=1.00$ GeV
will be similar to the result of the MCM
if the same parametrizations for the strange (anti-strange) distributions
are employed in the two models.
Indeed our following numerical calculations confirm this expectation.

To make more concrete
our above discussions about the fluctuation functions, we employ
the same parametrizations for the strange and anti-strange distribution
in the $\Lambda$ ($\Sigma$) and $K$ i.e. Eq.~(\ref{sB}) for the $s^{B}$
and Eq.~(\ref{sbarK}) for the $\sbar^K$, in the two models.
The LCM in this case is denoted as modified light-cone model (MLCM).
We present the numerical results for different values of
$\alpha$ and $\Lambda=1.08$ GeV in Fig.~7.
It can be found that
the prediction from the MLCM depends strongly on the value of $\alpha$
-- the calculation with $\alpha=0.33$ GeV is quite different from the
result of MCM while the calculation with $\alpha=1.00$ GeV is quite similar
to the result of MCM.
Changing the value of $\Lambda$, which corresponds to changing the possibilities
of finding the baryon-meson Fock state in the nucleon, will not greatly affect
our above conclusion.

\section{Summary}

The $s$-$\sbar$ asymmetry in the nucleon sea is an important observable
in the understanding of nucleon structure and the strong interaction.
Some theoretical attempts have been made in this direction.
Among them the meson cloud model and the light-cone model give
significantly different predictions
for both the shape of $s(x)-\sbar(x)$ and the absolute magnitude.
By reanalyzing these two models carefully, we point out that
although the quantitative calculations in the two models both depend strongly on
the model parameters, the predictions for the shape of $s(x)-\sbar(x)$
from the two models are parameter independent and opposite.
Thus it should be easy for more precise experimental data
to distinguish between the two models. 
The reasons for the quite different predictions obtained in the two models
%although the same nucleon fluctuating to baryon and $K$ meson mechanism
%and two-level convolution formula are used
are that the strange and anti-strange distributions in the baryon and meson,
and the fluctuation functions employed in the two models
are dramatically different.

To further investigate the two models, we modify the LCM by employing
the same parametrizations for the strange and anti-strange quark distributions
in the $\Lambda$ ($\Sigma$) baryon and $K$ meson in the LCM as that in the MCM.
It is found that the calculation in the modified light-cone model (MLCM)
depends strikingly on the parameter $\alpha$ which controls
the behavior of the meson-baryon fluctuation function.
Consequently, the $s$-${\bar s}$ asymmetry of the
nucleon sea from the MLCM can be different dramatically
from that from the MCM in some parameter ranges,
as well as very similar in other parameter ranges.
The coming experimental data will examine these calculations,
or equally, provide information on both the fluctuation function and
strange distributions in the $\Lambda$ ($\Sigma$) baryon and $K$ meson.

\section*{Acknowledgments}
This work was partially supported by the Massey Postdoctoral Foundation, New Zealand.

%\newpage

\newpage
%\vskip 1cm
\section*{Figure Captions}
\begin{description}
\item
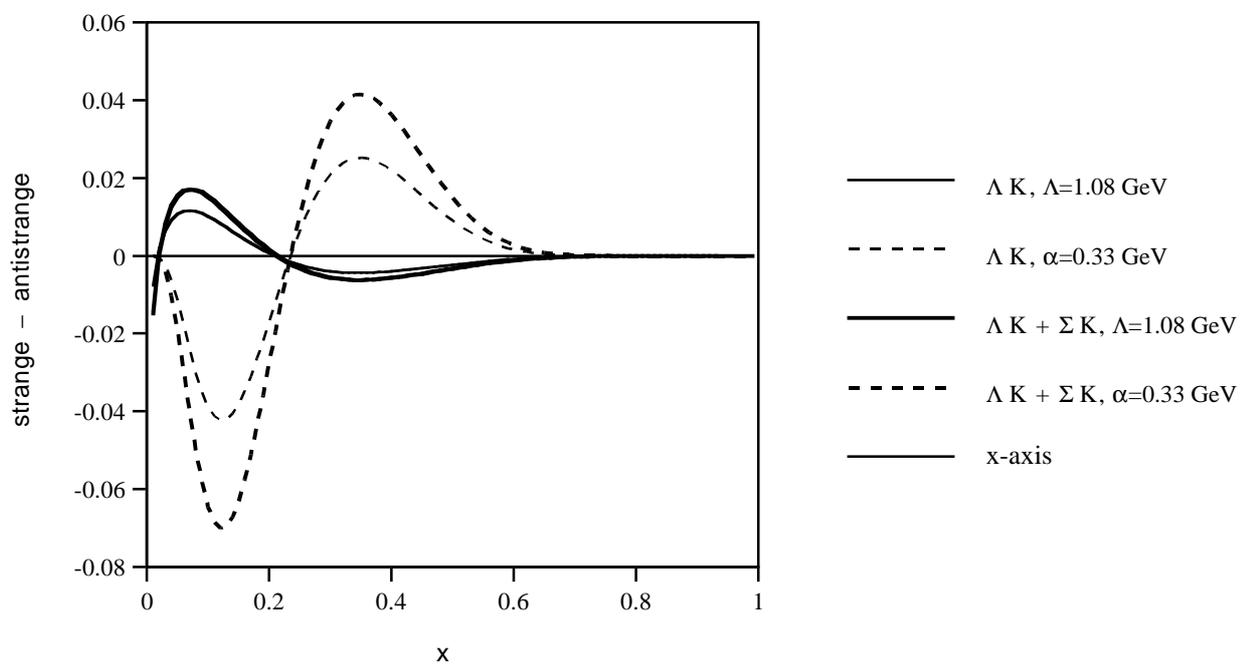
{Fig.~1.} 
$s(x)-\sbar(x)$ calculated in the MCM and LCM.
The thin solid and thick solid curves are
the contributions from $\Lambda K$ component
and $\Lambda K$ plus $\Sigma K$ components respectively,
obtained in the MCM.
The cut-off parameter in the form factor involved in the MCM
has been taken as $\Lambda=1.08$ GeV.
The thin dashed and thick dashed curves are
the contributions from $\Lambda K$ component
and $\Lambda K$ plus $\Sigma K$ components respectively,
obtained in the LCM with the parameter $\alpha=0.33$ GeV.
\item
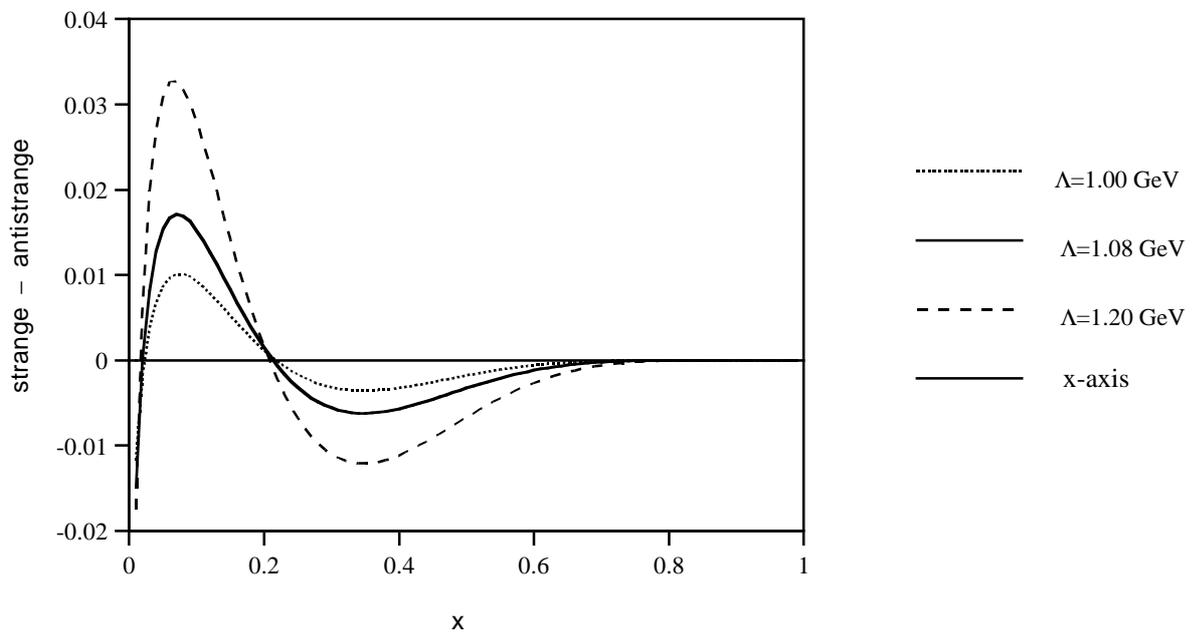
{Fig.~2.}
$s(x)-\sbar(x)$ calculated in the MCM with
both $\Lambda K$ and $\Sigma K$ components included.
The dotted, solid, and dashed curves are the results
with cut-off parameters $\Lambda=1.00$~GeV, $1.08$~GeV, and
$1.20$~GeV respectively.
\item
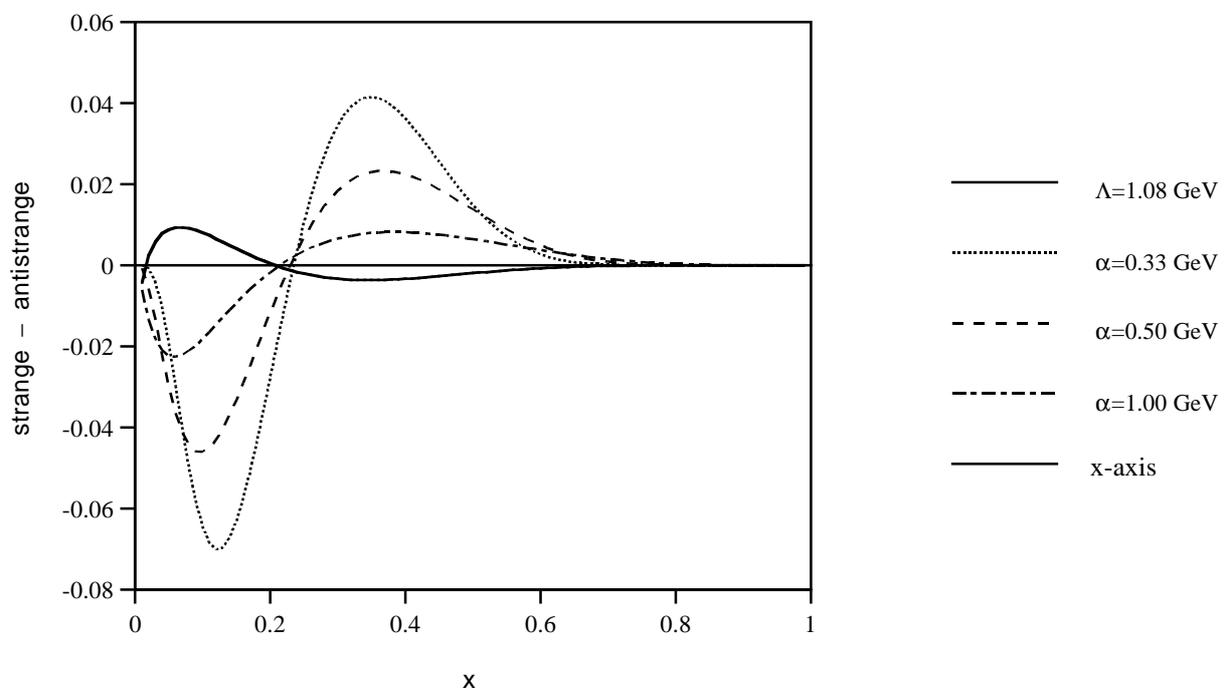
{Fig.~3.}
$s(x)-\sbar(x)$ calculated in the MCM and LCM with
both $\Lambda K$ and $\Sigma K$ components included.
The solid curve is the result from the MCM with $\Lambda=1.08$ GeV.
The dotted, dashed, and dashed-dotted curves
are the predictions from the LCM with
$\alpha=0.33$~GeV, $0.50$~GeV and $1.00$~GeV respectively.
\item
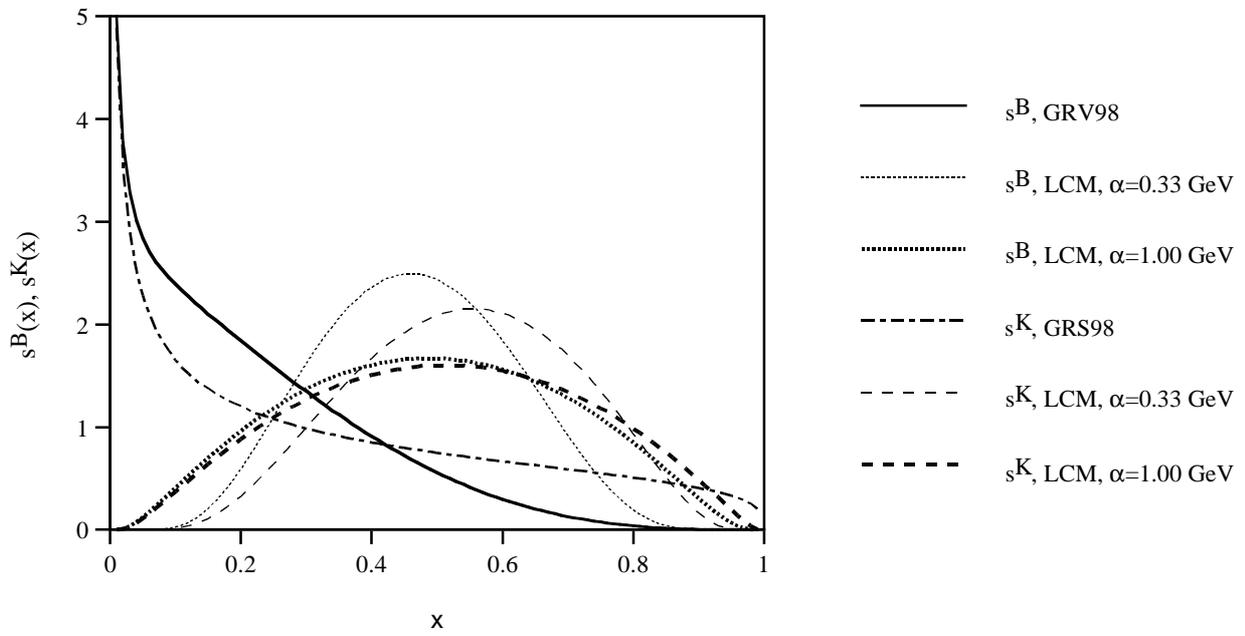
{Fig.~4.}
$s$ and $\sbar$ distributions in the baryon $\Lambda$ ($\Sigma$)
and the meson $K$.
The solid curve is the parametrization of GRV98 for $s^B$.
The thin dotted and thick dotted curves are the $s^B$ employed
in the LCM with $\alpha=0.33$~GeV and $1.00$~GeV respectively.
The dashed-dotted curve is the parametrization of GRS98 for ${\sbar}^K$.
The thin dashed and thick dashed curves are the ${\sbar}^K$ employed
in the LCM with $\alpha=0.33$~GeV and $1.00$~GeV respectively.
\item
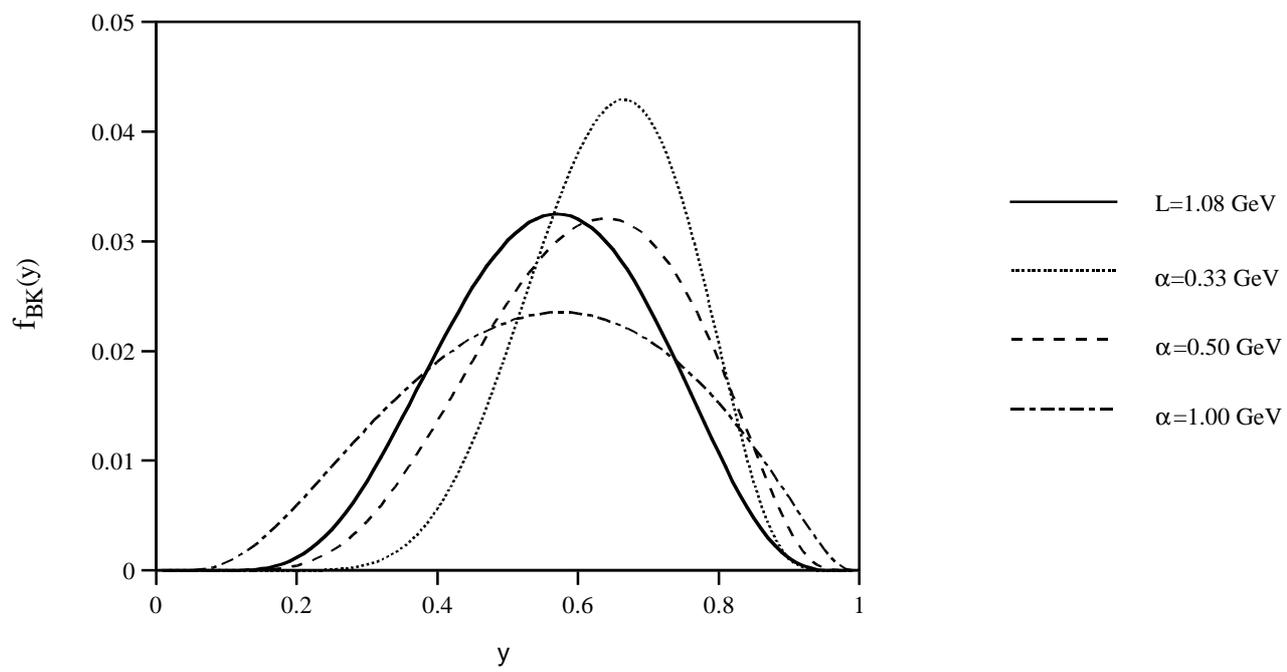
{Fig.~5.}
Fluctuation functions for $N \rightarrow \Lambda K$
with the probability being $1.27\%$.
The solid curve is the result from MCM with $\Lambda=1.08$~GeV.
The dotted, dashed, and dashed-dotted curves are the results from LCM with
$\alpha=0.33$~GeV, $0.50$~GeV, and $1.00$~GeV respectively.
\item
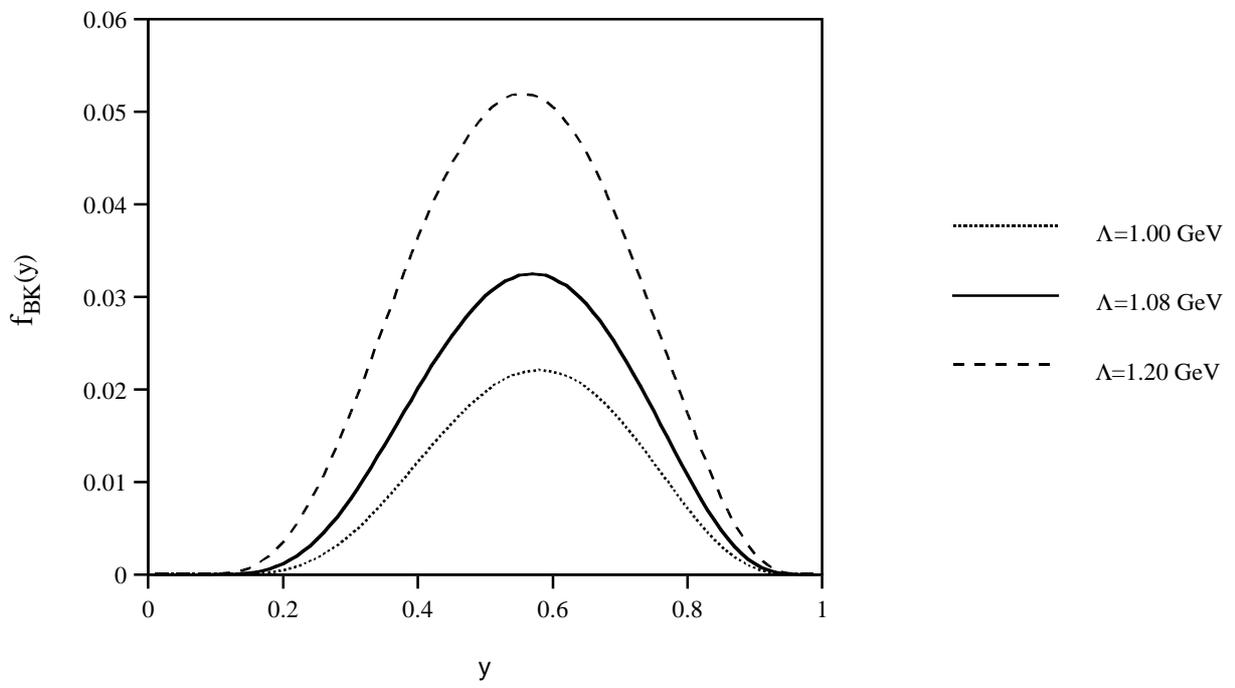
{Fig.~6.}
Fluctuation functions for $N \rightarrow \Lambda K$
calculated in the MCM.
The dotted, solid, and dashed curves are obtained with
$\Lambda=1.00$~GeV, $1.08$~GeV, and $1.20$~GeV respectively,
which corresponds to $P_{\Lambda K}=0.83\%, \, 1.27\%, \, 2.42\%$ respectively. 
\item
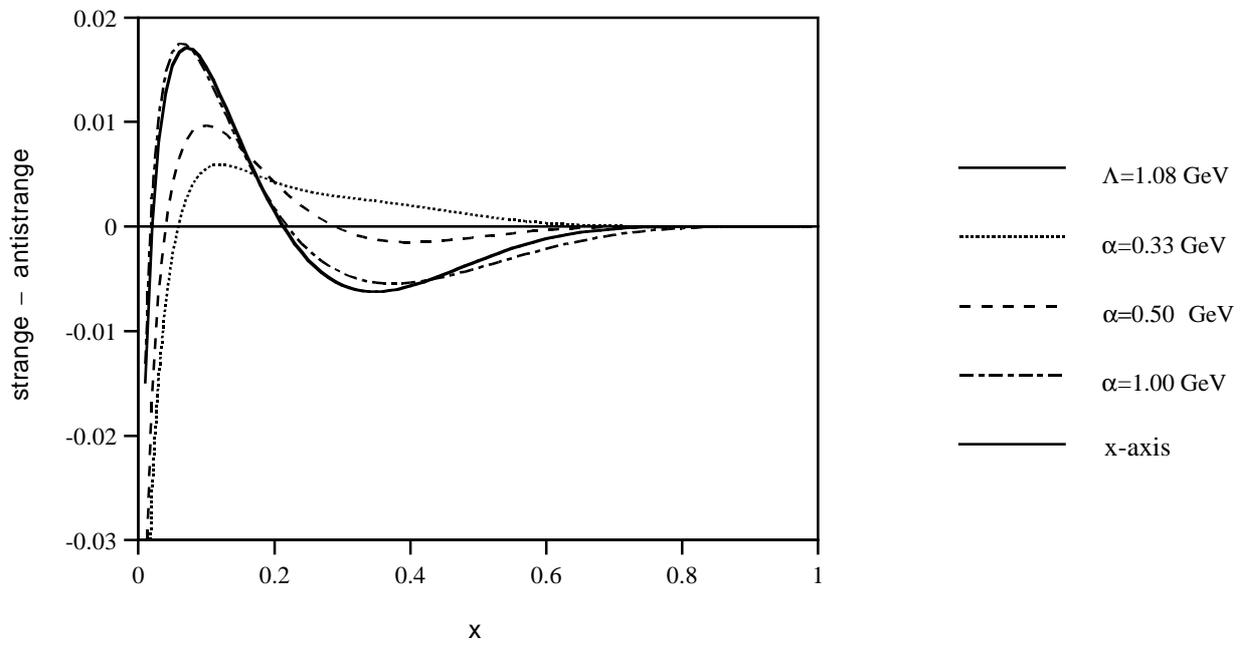
{Fig.~7.}
$s(x)-\sbar(x)$ calculated in the MCM and MLCM with
both $\Lambda K$ and $\Sigma K$ components included.
The solid curve is calculated in the MCM with $\Lambda=1.08$~GeV.
The dotted, dashed, and dashed-dotted curves are the results from MLCM with
$\alpha=0.33$~GeV, $0.50$~GeV, and $1.00$~GeV, respectively.
\end{description}

\end{document}